# Ergodicity breaking of an inorganic glass in aging near $T_g$ probed by elasticity relaxation

Wang Jianbiao, Wang Xu, and Ruan Haihui *
*Department of Mechanical Engineering, The Hong Kong Polytechnic University, Hung Hum, Kowloon, Hong Kong, China*
*e-mail address: haihui.ruan@polyu.edu.hk

We performed a series of aging experiments of an inorganic glass ($As_2Se_3$) at a temperature $T_2$ near the glass transition point $T_g$ by first relaxing it at $T_1$. The relaxation of Young's modulus was monitored, which was (almost if not ideally) exponential with a $T_1$-dependent relaxation time $\tau(T_1)$. We demostrate the Kovacs' paradox for the first time in inorganic glasses. Associated with the divergence of $\tau$, the quasi-equilibrated Young's modulus $E_\infty$ does not converge either. An elastic model of relaxation time and a Mori-Tanaka analysis of $E_\infty$ lead to a similar estimate of the persistent memory of the history, illuminating the ergodicity breaking within the accessible experimental time. Experiments with different $T_2$ exhibits a critical temperature $T_p \sim T_g$, *i.e.*, when $T_2 > T_p$, both $\tau$ and $E_\infty$ converge.

When a glass is aged at a temperature for which the decay of a physical variable is measurable, one intuitively expects that the dependence on the initial state, or the memory, should decay to zero based on the ansatz that the equilibrium state of supercooled liquid is approachable and unique [1]. That is, a glassy system is still ergodic at sufficiently long times even below the glass transition point, $T_g$, as the latter is merely an inflection point of a continuous cooling/heating curve, which does not indicate any critical feature of symmetry breaking. The ultimate fate of being ergodic is the cornerstone of the constitutive theories of glass [2], *e.g.*, the Tool-Narayanaswamy-Moynihan (TNM) [3-5] and the Kovacs-Aklonis-Hutchinson-Ramos (KAHR) models [6], where a fictive temperature evolving towards the bath temperature is hypothesized to account for the fading memory. However, this ansatz was challenged by Kovacs' experimental study on the volume relaxation of a polymeric glass below $T_g$ in 1964 [7], showing that the effective relaxation time ($\tau_{eff}$) after long-time aging was still affected by the initial state especially in the experiments of temperature up-jump. Because of the presumed uniqueness of an equilibrium (ergodic) state (note that the equilibrated volume was not provided by Kovacs [7]), the observed divergence of $\tau_{eff}$ was termed as "expansion-gap paradox" or "$\tau_{eff}$ paradox" [2, 8], questioning how a *quasi*-equilibrated system can exhibit disparate dynamics. Owing to the fundamentality in understanding glass relaxation, Kovacs' experiment was later re-examined by McKenna *et al* [2] and repeated by Koll and Simon [8], which confirmed the gap in the *quasi*-equilibrium $\tau_{eff}$.

In other glassy systems such as a charge-density glass [9] and more extensively studied spin glasses [10], aging experiments also revealed a non-vanishing dependence on historic disturbance especially when the latter was applied with a long waiting time. This was manifested by not only the varied spectrum of relaxation time [10], corresponding to Kovacs' finding, but also the non-converged physical quantity in an experimentally accessible duration [11]. Such observations together with the similar results of numerical simulations [12, 13] have led to the concept of ergodicity breaking (EB) described by a phenomenological model based on a rugged free-energy surface [13], or more analytically, replica symmetry breaking (RSB), revealed in the mean-field solutions of spin systems [14, 15]. Therefrom, to our understanding, a spin-glass transition, occurring at a critical temperature $T_C$, can be precisely defined (*i.e.*, weakly dependent on cooling rate [13]), as it manifests EB. In aging, it signifies a phenomenon that the long-term memory of historical disturbance persists below $T_C$, and *vice versa* vanishes when $T > T_C$ [10, 13]. While it is generally believed that the results of spin-glass models can be extended to structural glasses, it is still arguable on the possibility of finding any EB phenomenon in the latter, considering that the built-in randomness of spin-spin interactions differs fundamentally from the self-generated position randomness of structural glasses.

Kovacs' finding of the expansion gap seemingly hints at this possibility. However, his results were criticized especially by the inorganic glass community [16-18] as the dilatometry experiments with inorganic glasses [19, 20] after Kovacs did not render a convincing trend of persistent history dependence near $T_g$. For example, Goldstein and Nakonecznyj [19] speculated that $\tau_{eff}$ paradox might not be found in inorganic glasses because they had a narrower relaxation-time spectrum than a polymeric glass did. Struik [21] criticized that the $\tau_{eff}$ paradox might merely be a manifestation of the divergence of $\tau_{eff}$ when a stretched exponential process approached equilibrium. These criticisms might have discouraged the effort following the route of Kovacs to unveil an EB phenomenon or the nature of glass transition in structural glasses through monitoring volume ($V$) change.

In this work, we switched our attention to the variation of Young's modulus of a structural glass because it is a two-time quantity (*i.e.*, the autocorrelation function of stress or strain [22, 23]) that can be analogous to the magnetic susceptibility of a spin-glass and must be more sensitive to the heterogeneous dynamics in a glass [24]. Also, in experiments with structural glasses, $E$ changes much more significantly in the temperature range of glass transition. Roughly speaking, -lg(d$E$/d$T$/$E$) is 2 – 3 [25, 26] and -lg(d$V$/d$T$/$V$) is 5 – 6 [25] for an inorganic glass, *i.e.*, the variation in Young's modulus is at least two orders of magnitude more significant than that in volume at temperatures near $T_g$. Hence, it is more plausible to probe an EB phenomenon, if any, based on $E(t)$ than that based on $V(t)$. Hereunder, we reported aging experiments with an inorganic glass, in which the relaxations of $E$ were exponential. The exponentiality thus allows an undoubtful quantification of relaxation time $\tau$ as well as a clear quasi-equilibrium magnitude of $E_\infty$ when the aging time $t$ is much larger than $\tau$. Most unexpectedly, we show that both $\tau$ and $E_\infty$ depend identically on thermal history, *i.e.*, they do not converge at low temperatures but converge at high temperatures. These results infer a zero-to-nonzero transition at a critical point in the glass transition range.

The instantaneous Young's modulus was measured using the impulse excitation technique (IET), a nondestructive modulus measurement approach based on free vibration [26]. With a dedicated IET furnace, this technique has been employed to study relaxation phenomena in various glasses [25-29] as well as the transient amorphous states during a crystalline phase transition [30]. In the experiment, the sample was first heated to and annealed at an initial temperature $T_1$ for a sufficiently long time to reach a *quasi*-equilibrium state (no apparent trend of modulus change) and then quickly (about 0.5 °C/s) heated or cooled to the final temperature $T_2$. We chose[1] the commercial chalcogenide glass As$_2$Se$_3$ (Hubei New Hua-Guang Information Materials Co., Ltd, China) with the dilatometry $T_g$ = 180 °C, at which the viscosity is approximately $10^{12}$ Pa·s [31]. As$_2$Se$_3$ glass is a representative of chalcogenide glasses which has been widely used in infrared imaging [32] and optical switches [33]. It has excellent thermal stability against crystallization [34] with the lowest crystallization temperature of 200 °C, as extrapolated in the plot of isothermal crystallization rate [35]. The results presented in this letter were obtained from a sample of 40.025×8.035×2.45 mm$^3$ and 3.6396 g, measured using an IET station HT1600 (IMCE, Belgium). During heating, argon gas was purged to protect the sample from oxidation and Young's moduli were measured and recorded every 20 seconds. Fig. 1 shows the results of aging at $T_2$ = 175 °C after the temperature jumps from $T_1 = T_2 \pm \Delta T$ with $\Delta T$ = 5, 10, and 15 °C. An example temperature profile of the two-step aging is shown in the above-left inset, which illustrates the temperature overshoot and slow variations in a jump from $T_1$ =160°C to $T_2$=175°C. Though this transition took hundreds of seconds, it is regarded to be a short transient process in comparison with the long aging time at $T_1$ and $T_2$ (~ $10^4$ s). Also, the relaxation at $T_2$ starting from $t$ = 0 as defined in this inset, which warrants the measurements of quasi-equilibrium relaxation time and Young's modulus.

As shown in the main plot of Fig. 1, when $\Delta T$ = 5 °C, the relaxation processes after up and down jumps are almost symmetric and the Young's modulus after a long relaxation seemingly merges. With the increase in $\Delta T$, the relaxations after up and down jumps become asymmetric, and more surprisingly, the quasi-equilibrium Young's moduli are also different. We averaged the final leveled segment of the Young's modulus data, containing over 400 data points collected in hours, to quantify the *quasi*-equilibrium magnitude of Young's modulus $E_\infty$, as plotted in the above-right inset of Fig. 1 against $T_1$. Though the difference of $E_\infty$ is small (< 1.5%), it is noteworthy that the divergence of $E_\infty$ at $T_2$ is systematic, *i.e.*, $E_\infty$ decreases with $T_1$ and up-jump experiments render more deviation than down-jump ones when $\Delta T$ is the same.

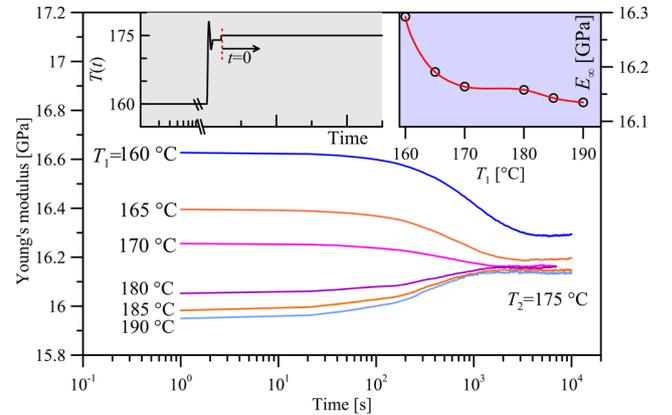

Fig. 1 Two-step aging results of As$_2$Se$_3$ glass with $T_2$ = 175 °C and $T_1 = T_2$ ± 5, 10, and 15 °C. Main plot: relaxation and *quasi*-equilibrated Young's modulus at $T_2$, showing that the divergence of Young's modulus relaxation time process, illustrating a $T_1$-dependence. The above-left inset: a temperature profile from 160 to 175 °C; the above-right inset: plot of quasi-equilibrium Young's modulus $E_\infty$ against $T_1$.

Using Kovacs's defination [7], we study the normalized modulus change $\delta_E(t) = (E(t)-E_\infty)/E_\infty$, as shown in Fig. 2(a) with the fitting curves using a stretched/compressed exponential function:

---

[1] Several oxide glasses were attempted. But the results were much more contaminated by experimental fluctuations, causing ambiguity to make any judgement. A plausible cause could be the noise in high-temperature measurements because of the higher $T_g$ (~ 500 °C). The IET system we employed is to record the sound generated by sample vibration. At a higher temperature, the environmental noise more deteriorates the weak acoustic signal.

$$\delta_E(t) = \delta_0 \exp\left[-(t/\tau)^\beta\right], \qquad (1)$$

where $\tau$ is the relaxation time, $\beta$ the stretching/compressing exponent, and $\delta_0$ a scaling constant. As shown in Fig. 2(a), all the relaxation curves are well fitted, and the fitting parameters of $\tau$ and $\beta$ are plotted in Figs. 2(b) and (c) against the initial temperature $T_1$. We notice that $\beta$ is almost unity with a maximum deviation of 0.07. Therefore, it is safe for us to claim that the relaxation $E(t)$ in the chalcogenide glass As$_2$Se$_3$ is exponential with negligible stretching and that Kovacs' finding of the non-converged relaxation time has been explicitly shown in Fig. 2(b); more specifically, $\tau$ decreases with the increase of $T_1$.

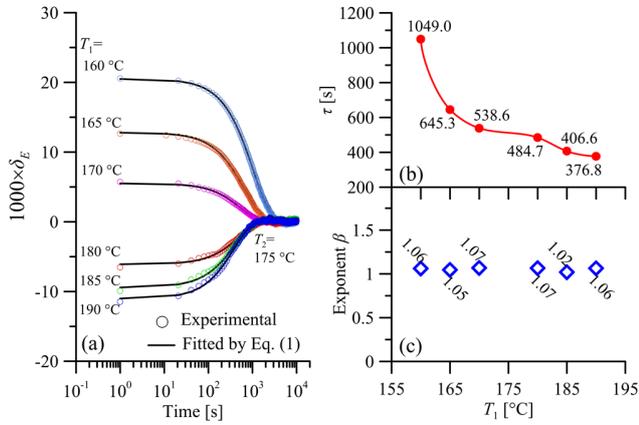

Fig. 2 (a) Plots of the relative change of Young's modulus $\delta_E(t)$ with fitting results using Eq. (1), and plots of (b) relaxation time $\tau$ and (c) exponent $\beta$ against $T_1$, indicating exponential relaxations

While various models have been proposed to reconcile the conflict between equilibrium dynamics and the history dependence indicated by Kovacs' $\tau_{\text{eff}}$ paradox, such as the rational thermodynamics [23], stochastics relaxation models [26-28], or the coupling model [36], a loophole in the paradox is indeed the experimental inability to probe the slowest relaxation in a non-exponential process [8, 21]. In our experiments, however, the relaxations are exponential, the loophole vanishes, and the history dependence of the relaxation time stands. Furthermore, we have supplemented the observation that the long-term relaxation may not bring the system to equilibrium because $E_\infty$ does not converge, i.e., the ergodicity is broken and the glassy system can only explore a $T_1$-dependent subregion of the configurational space. This finding echoes the extensive computational and experimental findings that the aging of a spin glass system, at temperatures below $T_g$, does not bring the system asymptotically to an equilibrium state [11-13]. Note that we cannot exclude the possibility that an ultra-long relaxation may bring the system to equilibrium because of the constraint of the experimental system (in our case it is due to the limit of inert gas supply). However, because of all the $E$–$t$ curves clearly leveled off as shown in Fig. 1 with durations of the flat segments over $10\tau$, it is reasonable to claim that the further relaxation, if exists, needs a timescale well beyond experimentally accessible range.

We anticipate further a correlation between the dynamics, manifested by $\tau$, and the statics, manifested by $E_\infty$, because the temperature dependences of them display similar features. Based on the elastic model proposed by Mooney [37] which was derived based on Eyring's picture of local molecular movements [38], a relaxation comes about when thermal fluctuations generate a local expansion exceeding a certain critical value. Mooney [37] estimated the probability of these relaxation events interfered by the thermal longitudinal sound waves and proposed that:

$$\tau = \tau_0 \exp\left[\frac{Q}{k_B T}\right], \qquad (2)$$

where $\tau_0$ is a pre-factor, $k_B$ is the Boltzmann constant, $Q \propto c_\infty^2 \propto E$ is the activation energy, and $c_\infty$ is the speed of longitudinal sound waves. Note that the temperature $T$ in Eq. (2) is a phonon temperature which does not account for the effect of the non-equilibrium dynamics associated with an unrelaxed atomic configuration [39]. We follow Tools [3] and other researchers [40, 41] to involve the effect of structural temperature by introducing an equivalent temperature $T_e \in [T_1, T_2]$ to replace $T$ in Eq. (3), because the glass is sufficiently equilibrated at $T_1$ and then aged at $T_2$. For simplicity, letting $T_e = \mu T_1 + (1-\mu)T_2$ with $\mu = \mu(t) \in [0, 1]$ being time-dependent, Eq. (2) is then recast as

$$\ln \tau = \ln \tau_0 + h\frac{E}{T_e}, \qquad (3)$$

where $h = Q/(k_B E)$. At a quasi-equilibrium state, $E_\infty$ and $\mu_\infty = \mu(t \to \infty)$ are constant. Replacing $E$ with $E_\infty$ in Eq. (3), and fitting the results of $\tau$, it is obtained that $\mu_\infty = 0.080$, as shown in Fig. 3, wherein the data points collapse to a straight line given by Eq. (3). The obtained slope $h = 1877.3$ K/GPa, together with $E_\infty = 16.2 \pm 0.1$ GPa (see the above-right inset of Fig. 1), leads to the activation energy of $Q = hEk_B = 60.4 \pm 0.4$ kcal/mol, agreeing reasonably well with the activation energy of 68 kcal/mol of As$_2$Se$_3$ near $T_g$ [42] determined based on the temperature dependence of shear viscosity.

The small value of $\mu_\infty$ indicates the non-vanishing structural memory in an aged glass, corresponding to the unmerged $E_\infty$ as shown in Fig. 1(b). We now draw a simplified picture for the aged glass to be a composite, wherein the matrix is a fully equilibrated (no memory) system at $T_2$ and inclusions are the persistent structure of $T_1$, as sketched in the inset of Fig. 3(b). Denoted by $E_i$ and $E_m$ the Young's moduli of the inclusions and the matrix, respectively, the ratio $E_\infty/E_i$ can be expressed as a function of $x = E_i/E_m$. Based on the Mori-Tanaka (MT) method by neglecting the shape of inclusions (i.e., assuming spherical inclusion), it is derived that $E_\infty/E_i = f(x)|_{(v_i, v_m, V_f)}$, *i.e.*, a function of $x$ with parameters $v_i$, $v_m$, and $V_f$ being the Poisson's ratio of inclusions, that of the matrix, and the volume fraction of inclusions (see Supplementary for details). Interestingly, $f(x)$ leads to $V_f$ because the function is very weakly dependent on $v_i$ and $v_m$. $E_m$ can be determined to be the average of $E_\infty$ of the up and down jumps with $\Delta T$ = 5°C as the two magnitudes are very close. Note that $E_i$ corresponds the Young's modulus of the *quasi*-equilibrium structures of $T_1$ quenched at $T_2$, which can be determined based on the Debye-Grüneisen coefficient (see Supplementary for details). The calculated experimental points of $f(x)$ are plotted in Fig. 3(b) together with the well fitted curve of MT theory with parametres of $v_i = v_m = 0.3$ and $V_f = 0.08095$. If $v_i$ and $v_m$ vary between [0.1, 0.4], $V_f$ varies between [0.08094, 0.08101] based on the best fit of experimental points, exhibiting a very weak dependence on Poisson's ratios. Intriguingly, $V_f$ is identical to $\mu_\infty$ though they are determined in completely different ways. This may imply that the quasi-equilibrium dynamics and atomic arrangement are closely correlated.

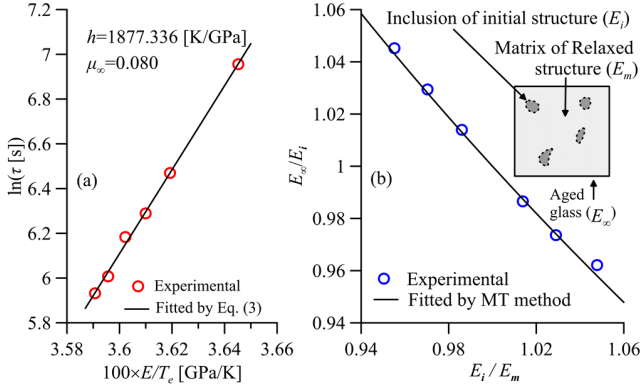

Fig. 3 Analysis of $T_1$ dependence of (a) relaxation time $\tau$ and (b) quasi-equilibrium Young's modulus $E_\infty$ based on the elastic model [37] and Mori-Tanaka approach, respectively, with an inset in (b) illustrating the simplification of the $T_2$-aged glass to be a composite

Noting that $[df(x)/dx]|_{(x=1, v_i=v_m)} + 1 = V_f$, *i.e.*, the slope in the plot of Fig. 3(b) near $x = 1$ can be used to estimate $V_f$. We repeated the aging experiments with $T_2$=170, 173, 177 and 180°C, and $T_1 = T_2 \pm 10$°C to obtain the relation between $V_f$ and $T_2$, i.e., the temperature dependence of memory persistence. The results are plotted in Fig. 4 (see Supplementary for details). To exemplify, the relaxation curves of $T_2$=170°C and 180°C are shown in the insets of Fig. 4. We notice that at 180 °C the two curves merge after a long-time aging and the exponential relaxation times after up and down jumps are almost identical (~ 300 s). Therefore, $V_f$ = 0 at 180°C. Noteworthily, when $T_2 = 177$°C, $V_f$ is 0.005 determined from the separate values of $E_\infty$ after up and down jumps. Besides, $\tau$ = 394.0s and 320.2s based on exponential fits for up and down jumps at this temperature, respectively, with the expected difference that the up-jump case relaxed slower. Fig. 4 suggests there is a critical temperature $T_p \in$ (177°C, 180°C) that can be analogous to $T_C$ of a spin glass. Below $T_p$, the structural memory persists, *i.e.*, ergodicity is broken; above it, the memory can fade completely, *i.e.*, the system restores ergodicity. For As$_2$Se$_3$, $T_g$ is not a uniquely determined temperature but varies in the range of 175 – 180 °C [43] due to the variety of characterization techniques. We hence argue that the critical temperature $T_p$ is within the empirical range of $T_g$, at least for As$_2$Se$_3$, signifying that structural glass transition is not just a slowing-down process. Also, we emphasize that our work has paved the way to uniquely determine $T_p$ through measuring $E_\infty$ after two-step aging with $\Delta T \sim 10$°C. The measurement of the exact $T_p$ for As$_2$Se$_3$ would only depend on the accuracy and resolution of temperature control and modulus measurement.

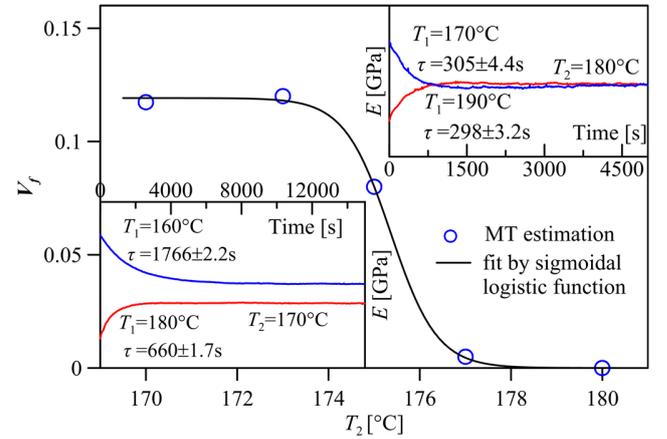

Fig. 4 Plot of the measure of memory persistence $V_f$ against the aging temperature $T_2$, indicating a clear transition at a critical temperature $T_C \in$ (177 °C, 180 °C) with bottom-left and upright insets showing the aging curves $E(t)$ at $T_2$ = 170 and 180 °C, respectively, after equilibrated at $T_1 = T_2 \pm 10$°C.

Phenomenologically, the persistent memory can be explained based on a rugged free energy landscape [13], namely, a glassy system may be trapped in deep energy basins during aging at $T_1$, constraining the exploration of the full energy landscape at $T_2$ within an experimentally accessible time that is already much longer than the relaxation time estimated from viscosity. This picture is also reminiscent of the "mosaic" transition delineated by the random first-order transition theory [44], that is, a glassy system transforms into a patchwork consisting of distinguishable atomic arrangements below $T_g$. In the two-step aging experiments, some patches at $T_1$ persists at $T_2$ during the long aging.

However, the mean-field picture based on a free-energy landscape does not explain how a persistent memory forms from the random variations of atomic configurations, especially when $T_p \sim T_g > T_K$ ($T_K$, the Kauzmann temperature).

Therefore, a real-space picture is needed. Trying to establish it, at least with a clue, we note several recent attempts in establishing the connection between static structures and long-time dynamics based on molecular dynamics (MD) simulations. In a very recent investigation of structural-property relation in a binary Lennard–Jones glass, a machine learning algorithm [45] was established, which, after training, can predict long-time dynamics based on the static initial structures. This result is a triumph in decoding the nature of glass state, as it unveils that the information of initial structures is not "forgotten" in the subsequent relaxation process [46], similar to our results on the initial-state dependence of relaxation time. More transparently, Wang et al. [47] conducted MD simulations of $Cu_{50}Zr_{50}$ and found that the activation energy had a strong correlation with the vibrational mean squared displacement (VMSD) instead of the short-range structural indices. As VMSD describes the long-range elastic interactions, Wang et al. [47] argued that the effect of elastic constraint could be the rate-limiting mechanism of structural relaxation in a glass, which is also the reason we employed the elastic model to analyze relaxation time.

Interestingly, Wang et al. [47] found that the glass tends to be more heterogeneous during relaxation because the soft spots, those substructures with higher flexibility (higher free energy), tended to flock together, leading the heterogeneity in both structure and dynamics. Parallelly, Zhang and Lam [48] have established a distinguishable-particle lattice model (DPLM), which can be regarded as an abstraction of soft-spot dynamics. It simplifies position randomness in a structural glass to be a random force field between site particles and introduces voids to mimic the motion of soft spots. Such a setup leads to the *spatially constrained dynamics* (SCD), *i.e.*, only agminated voids bring about significant relaxation events while isolated voids are trapped. The DPLM simulations successfully reproduced the divergence of *quasi*-equilibrium $\tau_{\text{eff}}$ [49] even though the void concentration and thus the equilibrium state were predefined. Encouraged by the DPLM result and MD simulations, we anticipate that a simplified real-space glass model revealing SCD may involve additionally the generation and depletion of soft spots (or voids) that is dependent on the bath as well as structural temperatures, local stress state, and global energy penalty, following the free-volume picture [50]. Thus, the initial-state dependence, in terms of both relaxation time and quasi-equilibrium state, may be revealed as a consequence of SCD and the associated evolution of the distribution of soft spots.

In conclusion, we performed two-step aging experiments with an inorganic glass $As_2Se_3$ and revealed for the first time the clear phenomena of ergodicity breaking based on the measurements of instantaneous Young's modulus. We identified a critical ergodicity-breaking temperature of $As_2Se_3$ in terms of the volume fraction of the persistent memory, which was within the empirical glass transition range.

*Acknowledgment* — This work was inspired by an interesting discussion with CH Lam and Matteo Lulli. We gratefully acknowledge their comments and the finical support of Hong Kong ECS (Grant No. 25200515) and GRF (Grant No. 15213619).

# Supplemental Material to "Ergodicity breaking of an inorganic glass in aging near $T_g$ probed by elasticity relaxation"

Jianbiao Wang, Xu Wang, and Haihui Ruan*

Department of Mechanical Engineering, The Hong Kong Polytechnic University, Hung Hum, Kowloon, Hong Kong, China

## I. Description of Mori-Tanaka (MT) method

By neglecting the shape of the inclusions, we can simply suppose the inclusion is spherical, and the effective shear and bulk modulus can be derived based on the Mori-Tanaka (MT) method [1]:

$$\begin{cases} G_\infty = G_m + \dfrac{(G_i - G_m)V_f}{1 + 4(1-V_f)G_p(G_i - G_r)} \\ K_\infty = K_m + \dfrac{(K_i - K_m)V_f}{1 + 9(1-V_f)K_p(K_i - K_m)} \end{cases}, \quad \text{(S1)}$$

where

$G_\infty$: effective shear modulus;
$G_i$: shear modulus of initial structures (inclusion);
$G_m$: shear modulus of relaxed structures (matrix);
$K_\infty$: effective bulk modulus;
$K_i$: bulk modulus of initial structures (inclusion);
$K_m$: bulk modulus of relaxed structures (matrix).

and

$$\begin{cases} G_p = \dfrac{3(2G_m + K_m)}{10G_m(4G_m + 3K_m)} \\ K_p = \dfrac{1}{3(4G_m + 3K_m)} \end{cases}. \quad \text{(S2)}$$

Recall the relations:

$$\begin{cases} G_\vartheta = \dfrac{E_\vartheta}{2(1+\nu_\vartheta)} \\ K_\vartheta = \dfrac{E_\vartheta}{3(1-2\nu_\vartheta)} \end{cases}, \quad \text{(S3)}$$

where $E_x$, and $\nu_x$ are Young's modulus and Poison's ratio, respectively; $\vartheta = i$ and $m$. Then the effective Young's modulus can be written as:

$$E_\infty = \dfrac{9K_\infty G_\infty}{3K_\infty + G_\infty} = g(E_i, E_r)\big|_{(\nu_i,\nu_m,V_f)}, \quad \text{(S4)}$$

That is, $E_\infty$ can be considered as a function of $E_i$ and $E_m$ with parameters of $(\nu_i, \nu_m, V_f)$. Normalizing Eq. (S4) with $E_i$ leads to

$$E_\infty / E_i = y(1, E_m/E_i)\big|_{(\nu_i,\nu_m,V_f)} = f(x)\big|_{(\nu_i,\nu_m,V_f)}, \quad \text{(S5)}$$

With some simple derivations, we obtain:

$$\dfrac{df(x)}{dx}\bigg|_{(x=1,\nu_i=\nu_m)} + 1 = V_f, \quad \text{(S6)}$$

## II. The quasi-equilibrium Young's moduli at $T_1$ and $T_2$

In Table S1, the quasi-equilibrium Young's moduli of $T_1$ are provided. The corresponding Young's moduli at $T_2$ have already provided in the main text. To calculate $V_f$ at varied $T_2$, the quasi-equilibrium Young's moduli of $T_1$ and $T_2$ are measured again. The results are provided in Table S2.

Table S1

| $T_1$ [°C] | $E_\infty$ [GPa] |
|---|---|
| 160 | 17.04 |
| 165 | 16.70 |
| 170 | 16.42 |
| 180 | 15.90 |
| 185 | 15.61 |
| 190 | 15.33 |

Table S2

| $T_1$ [°C] | $E_\infty$ [GPa] | $T_2$ [°C] | $E_\infty$ [GPa] |
|---|---|---|---|
| 160 | 17.01 | 170 | 16.51 |
| 180 | 15.88 | 170 | 16.39 |
| 163 | 16.84 | 173 | 16.77 |
| 183 | 15.82 | 173 | 16.28 |
| 167 | 16.56 | 177 | 16.052 |
| 187 | 15.52 | 177 | 16.046 |
| 170 | 16.39 | 180 | 15.88 |
| 190 | 15.34 | 180 | 15.88 |

As shown in Fig. S1, the Debye-Grüneisen coefficient $dE/dT$=0.0071 GPa/°C, based on the temperature-modulus measurements of As$_2$Se$_3$ glass at temperature range [40°C, 150°C] with the heating rate of 20 °C/min.

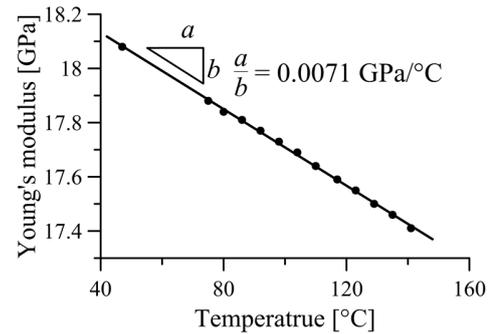

Fig. S1 The temperature dependence of Young's modulus of As$_2$Se$_3$ glass. The Debye-Grüneisen is estimated by a linear fit.